\documentclass[mypaper,7pt,twoside]{CoAst}
\usepackage{epsf,graphicx,fancyhdr}
\renewcommand{\chaptermark}[1]{\markboth{#1}{}}

\newcommand{\kopf}{\small\itshape Comm. in Asteroseismology\\ Vol. 150, 2007}
\newcommand{\Authors}[1]{\begin{center}\normalsize\bf\sf #1 \end{center}}

\renewcommand{\author}[1]{\begin{center}\normalsize\bf\sf #1 \end{center}}
\newcommand{\Address}[1]{\begin{center}\small\sf #1 \end{center}}

\newcommand{\References}[1]{\vspace{2.4mm}\begin{flushleft}{\large References\\}\vspace*{1mm}\small #1 \end{flushleft}}

\newcommand{\chapterDSSN}[2]{\chapter[\sf\normalsize #1\\ \footnotesize \hspace*{5mm}by #2 \sf\normalsize][]{#1\\}\rhead[\fancyplain{}{\sf\footnotesize \center{#1}}]{\fancyplain{}{\sffamily\thepage}}\lhead[\fancyplain{\kopf}{\sffamily\thepage}]{\fancyplain{\kopf}{\sf\footnotesize \center{#2}}}}

\newcommand{\figureDSSN}[5]{\begin{figure}[#4]
\centering
\includegraphics*[#5]{#1}
\caption{#2}
\label{#3}
\end{figure}}

\renewcommand{\chaptername}{}
\newcommand{\acknowledgments}[1]{\vspace*{5mm}\noindent\begin{bf}Acknowledgments. \end{bf} #1}

\begin{document}
\sf

\chapterDSSN{44 Tau: Discrimination between MS and post-MS models}
{P. Lenz, A. A. Pamyatnykh, M. Breger and V. Antoci}

\Authors{P. Lenz$^1$, A. A. Pamyatnykh$^{1,2,3}$, M. Breger$^1$ and V. Antoci$^1$}
\Address{$^1$ Institut f\"ur Astronomie, T\"urkenschanzstrasse 17, 1180 Vienna, Austria\\
$^2$ Copernicus Astronomical Center, Bartycka 18, 00-716 Warsaw, Poland\\
$^3$ Institute of Astronomy, Pyatnitskaya Str. 48, 109017 Moscow, Russia}

\section{Observations and Mode Identification}

Antoci et al. (2006) analyzed photometric data of 44 Tau from
2000-2003 and detected 29 oscillation frequencies of which 13
frequencies are independent. We performed a mode identification
based on the amplitude ratios and phase differences from the
photometric data set of 2000/01. As shown by
Daszy\'nska-Daszkiewicz et al.\,(2003), the results are very
sensitive to the treatment of convection in the envelope. We find
that in the case of 44 Tau only models with ineffective convection
($\alpha_{\rm conv} \approx 0$) result in definitive mode
identification. The observed modes f1 (6.8980 c/d) and f5
(8.9606 c/d) can definitely be identified as $\ell=0$ modes. 
Their frequency ratio 0.7698 is close to the typical ratio of the
radial fundamental and first overtone frequencies in the $\delta$
Scuti domain. Four modes are identified as $\ell=1$ and two other
modes as $\ell=2$ modes. Two $\ell=1$ modes (9.1175 and
9.5613 c/d) and both $\ell=2$ modes look to be close to the
avoided crossing stage and may be used as indicators of efficiency
of the overshooting from the stellar convective core.

\section{Modelling 44 Tau}

From the HIPPARCOS parallax, Str\"omgren and Geneva
photometry and from Vienna grid of stellar atmospheres (Nendwich et al., 2004) we derive
$\log L/L_{\odot}$ = $1.34 \pm 0.07$ and $T_{\mbox{\small eff}}$ =
$6900 \pm 100K$. With a $\log g$ value of $3.6 \pm 0.1$ it is not
possible to determine the evolutionary status of 44 Tau
unambiguously.

Our main sequence models of 44 Tau that can fit the radial
fundamental and first overtone generally are too cool and in some
cases too faint. An acceptable fit of all identified modes
can be obtained only for enhanced metal abundance and/or
significant overshooting from the convective core.

\figureDSSN{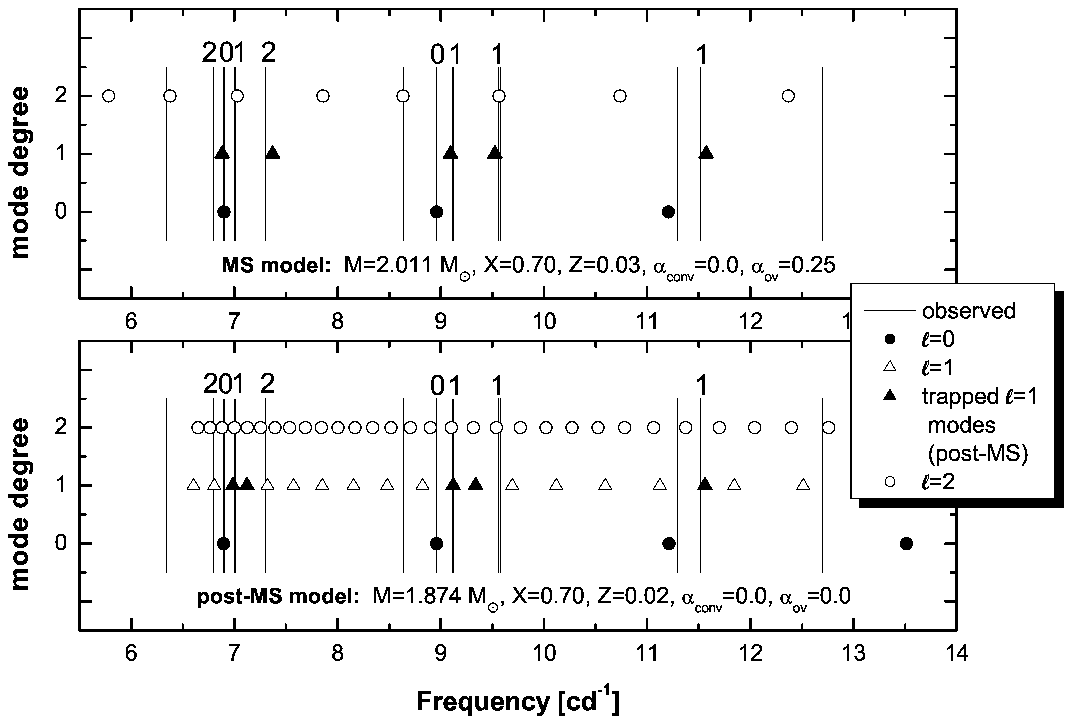}{Comparison of the predicted
frequency spectrum with observations for a selected main sequence
model (upper panel) and a post-MS model (lower panel). In
the post-MS case trapped $\ell=1$ modes are indicated by different
symbols.}{freqspec}{!ht}{clip,angle=0,width=105mm}

In the post-MS case it is possible to obtain a model within the
 observational error box in the HRD with no need of
overshooting and nonstandard chemical composition. For post-MS
models we predict much more unstable modes than we
observe. A possible explanation why only specific modes are
observed is mode trapping in the stellar envelope.

The predicted frequency spectra for the MS and post-MS case are given in Fig.~\ref{freqspec}.

\vspace{-2mm}
\section{Conclusions}

For both MS and post-MS models it is possible to obtain good fits
to the observed frequency spectrum. However, the MS models are
significantly cooler than it can be estimated from
photometry. Considering the good fit of both the observed
frequencies and physical parameters, standard post-main sequence
models with inefficient convection seem to be preferable.

\acknowledgments{ We would like to thank Rafa Garrido and Juan
Carlos Su\'arez for valuable discussions during the conference.
This work has been supported by the Austrian FWF. AAP
acknowledges the financial support from HELAS and from the Polish MNiI grant No. 1 P03D 021 28. }

\newpage
\References{
Antoci, V., Breger, M., Rodler, F., Bischof, K., Garrido, R. 2006, A\&A in press\\
Daszy\'nska-Daszkiewicz, J., Dziembowski, W. A., Pamyatnykh, A. A. 2003, A\&A 407, 999\\
Nendwich, J., Heiter, U., Kupka, F., et al. 2004, CoAst 144, 43
}

\end{document}